\documentclass[aps,print,showpacs,twocolumn]{revtex4}%
\usepackage{amsfonts}
\usepackage{amsmath}
\usepackage{amssymb}
\usepackage{graphicx}%
\providecommand{\U}[1]{\protect\rule{.1in}{.1in}}

\begin{document}
\title{The Peregrine rogue waves induced by interaction between the continuous wave
and soliton }
\author{Guangye Yang$^{1,2}$, Lu Li$^{1}$}
\email{llz@sxu.edu.cn}
\author{Suotang Jia$^{3}$}
\affiliation{$^{1}$Institute of Theoretical Physics, Shanxi University, Taiyuan, Shanxi,
030006, China}
\affiliation{$^{2}$Department of Physics, Shanxi Medical University, Taiyuan, Shanxi,
030001, China}
\affiliation{$^{3}$College of Physics and Electronics Engineering, Shanxi University,
Taiyuan 030006, China}

\pacs{42.65.Tg, 42.81.Dp, 42.79.Sz}

\begin{abstract}
Based on the soliton solution on a continuous wave background for an
integrable Hirota equation, the reduction mechanism and the characteristics of
the Peregrine rogue wave in the propagation of femtosecond pulses of optical
fiber are discussed. The results show that there exist two processes of the
formation of the Peregrine rogue wave: one is the localized process of the
continuous wave background, and the other is the reduction process of the
periodization of the bright soliton. The characteristics of the Peregrine
rogue wave are exhibited by strong temporal and spatial localization. Also,
various initial excitations of the Peregrine rogue wave are performed and the
results show that the Peregrine rogue wave can be excited by a small localized
(single peak) perturbation pulse of the continuous wave background, even for
the nonintegrable case. The numerical simulations show that the Peregrine
rogue wave is unstable. Finally, through a realistic example, the influence of
the self-frequency shift to the dynamics of the Peregrine rogue wave is
discussed. The results show that in the absence of the self-frequency shift,
the Peregrine rogue wave can split into several subpuslses; however, when the
self-frequency shift is considered, the Peregrine rogue wave no longer splits
and exhibits mainly a peak changing and an increasing evolution property of
the field amplitude.

\end{abstract}
\maketitle

\section{Introduction}

A rogue wave is an oceanic phenomenon with amplitude much higher than the
average wave crests around them \cite{Kharif_Book}. So far, this phenomenon
has not been understood completely due to the difficult and restricted
observational conditions. Therefore a great deal of attention has been paid to
better understanding their physical mechanisms. It has been suggested that the
rogue waves appearing in the ocean are mainly caused by wave-wave nonlinear
interaction, such as a modulation instability of the Benjamin-Feir type
\cite{Kharif,Ruban,Andonowati,Zakharov,Slunyaev}. Recently, rogue waves have
been observed in optical fibers \cite{Solli1}, in superfluid helium
\cite{Ganshin}, and in capillary waves \cite{Shats}, respectively. These
discoveries indicate that rogue waves may be rather universal. Certainly, we
cannot harness the occurrence of rogue waves in the ocean due to their
enormous destructiveness. In optics, however, the optical rogue waves produced
in supercontinuum generation can be used to generate highly energetic optical
pulses \cite{Solli1,Dudley,Mussot,Solli3,Taki}.

Deep waves in the ocean and the wave propagation in optical fibers can be
described by the nonlinear Schr\"{o}dinger (NLS) equation. Based on the model,
the rogue wave phenomenon has been extensively studied, including rational
solutions and their interactions
\cite{Osborne,Osborne2,Akhmediev1,Akhmediev2,Akhmediev3,Akhmediev4,Ankiewicz1,Yan}%
, pulse splitting induced by higher-order modulation instability, and wave
turbulence \cite{Erkintalo,Kibler}. A fundamental analytical solution on the
rogue waves is the Peregrine solution (PS), which was first presented by
Peregrine \cite{Peregrine}. PS is a localized solution in both time and space,
and is a limiting case of Kuznetsov-Ma solitons \cite{Kuznetsov,Ma} and
Akhmediev breathers \cite{Akhmediev8}. Recently, the excitation conditions of
PS have been demonstrated experimentally in optical fiber, and explicitly
characterized its two-dimensional localization \cite{NatPhys,Hammani}. It
should be noted that the results are theoretically described by the NLS
equation, which is valid for the picosecond pulses. When describing the
characteristics of PS in the femtosecond regime, we must consider some
higher-order effects, such as third-order dispersion (TOD), self-steepening
and self-frequency shift, and so on. In this case, we should consider the
higher-order nonlinear Schr\"{o}dinger (HNLS) equation in the form
\cite{Kodama}%
\begin{align}
\frac{\partial q}{\partial z}  &  =i\left(  \alpha_{1}\frac{\partial^{2}%
q}{\partial t^{2}}+\alpha_{2}\left\vert q\right\vert ^{2}q\right)  +\alpha
_{3}\frac{\partial^{3}q}{\partial t^{3}}\nonumber\\
&  +\alpha_{4}\frac{\partial\left\vert q\right\vert ^{2}q}{\partial t}%
+\alpha_{5}q\frac{\partial\left\vert q\right\vert ^{2}}{\partial t},
\label{He}%
\end{align}
where $q(z,t)$ is the slowly varying envelope of the electric field, $z$ and
$t$ denote normalized propagation distance and the retarded temporal
coordinate, respectively, and the parameters $\alpha_{1},\alpha_{2},\alpha
_{3},\alpha_{4},$ and $\alpha_{5}$ are the real constants related to the group
velocity dispersion (GVD), the self-phase modulation (SPM), the third-order
dispersion (TOD), the self-steepening and the delayed nonlinear response
effect, respectively.

Generally, the HNLS equation (\ref{He}) is not integrable. To solve Eq
(\ref{He}), we first consider the special parametric choice with $\alpha
_{2}=2\mu^{2}\alpha_{1}$, $\alpha_{4}=$ $6\mu^{2}\alpha_{3},$ $\alpha
_{4}+\alpha_{5}=0$ so that Eq. (\ref{He}) becomes \cite{Hirota}
\begin{equation}
\frac{\partial q}{\partial z}=i\left(  \alpha_{1}\frac{\partial^{2}q}{\partial
t^{2}}+\alpha_{2}\left\vert q\right\vert ^{2}q\right)  +\alpha_{3}%
\frac{\partial^{3}q}{\partial t^{3}}+\alpha_{4}\left\vert q\right\vert
^{2}\frac{\partial q}{\partial t}, \label{HirotaEq}%
\end{equation}
which is usually called the Hirota equation, where the parameter $\mu$ is a
real constant.

The equation was first presented by Hirota \cite{Hirota}, and subsequently
many researchers analyzed this equation from different points of view
\cite{Lakshmanan,Mihalache1,Mihalache2,Porsezian,Mihalache3,Li,Xu,lishuqing}.
Recently, the rogue waves and the rational solutions of the Hirota equation
have been discussed in the forms of the two lower-order solutions by employing
the Darboux transformation technique \cite{Ankiewicz2}. In this paper, based
on the soliton solution on a continuous wave (CW) background for the
integrable Hirota equation (\ref{HirotaEq}), we discuss the formation
mechanism and the characteristics of the PS in the femtosecond regime. The
results show that the PS exhibits a feature of temporal and spatial
localization and can be excited by a small localized (single peak)
perturbation pulse of the CW background, even for the nonintegrable HNLS
equation (\ref{He}). Based on this result, we discuss the dynamics of the
Peregrine rogue wave through a realistic example.

The paper is organized as follows. In Sec. II we present the explicit process
of the formation of the PS from the Kuznetsov-Ma solitons and Akhmediev
breathers based on the soliton solution on a CW background for the Hirota
equation and discuss the characteristics of the PS. Various initial
excitations of the PS are discussed in Sec. III. Subsequently, in Sec. IV we
investigate the influence of the self-frequency shift to the dynamics of the
Peregrine rogue wave by employing a realistic example. Our results are
summarized in Sec. V.

\section{The Peregrine solution induced by interaction between the continuous
wave background and soliton}

By employing a Darboux transformation, one can construct the soliton solution
on a CW background for Eq. (\ref{HirotaEq}) as follows \cite{Xu,lishuqing}%
\begin{align}
q(z,t)  &  =\left(  A+A_{s}\frac{a\cosh\theta+\cos\varphi}{\cosh\theta
+a\cos\varphi}\right. \nonumber\\
&  \text{ \ \ }\left.  +iA_{s}\frac{b\sinh\theta+c\sin\varphi}{\cosh
\theta+a\cos\varphi}\right)  \exp\left(  i\varphi_{c}\right)  . \label{S1}%
\end{align}
Here%
\begin{align*}
\theta &  =M_{I}T-\left(  \nu_{1}M_{R}+\nu_{2}M_{I}\right)  Z,\\
\varphi &  =M_{R}T-\left(  \nu_{2}M_{R}-\nu_{1}M_{I}\right)  Z,\\
\varphi_{c}  &  =\omega T+\left[  (2\mu^{2}A^{2}-\omega^{2})\alpha_{1}%
+\omega(6\mu^{2}A^{2}-\omega^{2})\alpha_{3}\right]  Z,
\end{align*}
with $T=t-t_{0}$, $Z=z-z_{0}$, and the coefficients are $a=-2\mu^{2}%
AA_{s}/(\mu^{2}A_{s}^{2}+M_{R}^{2})$, $b=-2\mu AM_{R}/(\mu^{2}A_{s}^{2}%
+M_{R}^{2})$, and $c=M_{I}/(\mu A_{s})$, which implies that $M_{I}=0$ as
$A_{s}=0$. And other coefficients $\nu_{1}=\mu A_{s}[\alpha_{1}+(\omega
+2\omega_{s})\alpha_{3}]$, $\nu_{2}=(\omega+\omega_{s})\alpha_{1}+(\omega
^{2}+\omega\omega_{s}+\omega_{s}^{2}-2\mu^{2}A^{2}-\mu^{2}A_{s}^{2})\alpha
_{3}$, $M_{R}=\{[((\omega_{s}-\omega)^{2}+\allowbreak4\mu^{2}A^{2}-\mu
^{2}A_{s}^{2})^{2}+4\mu^{2}A_{s}^{2}(\omega_{s}-\omega)^{2}]^{1/2}%
+((\omega_{s}-\omega)^{2}+4\mu^{2}A^{2}-\mu^{2}A_{s}^{2})\}^{1/2}/\sqrt{2}$,
and $M_{I}=\{[((\omega_{s}-\omega)^{2}+\allowbreak4\mu^{2}A^{2}-\mu^{2}%
A_{s}^{2})^{2}+4\mu^{2}A_{s}^{2}(\omega_{s}-\omega)^{2}]^{1/2}-((\omega
_{s}-\omega)^{2}+4\mu^{2}A^{2}-\mu^{2}A_{s}^{2})\}^{1/2}/\sqrt{2}$, and
$t_{0}$, $z_{0}$, $A$, $\omega$, $A_{s}$, and $\omega_{s}$ are the arbitrary
real constants, and without loss of generality we assume that $A$ and $A_{s}$
are non-negative constants. The solution (\ref{S1}) includes two special
cases. One is that as the amplitude $A$ vanishes, it reduces to the solution
$q_{s}(z,t)=A_{s}e^{i\varphi_{s}}\operatorname{sech}\theta$, where $\theta=\mu
A_{s}T-\mu A_{s}[2\omega_{s}\alpha_{1}+(3\omega_{s}^{2}-\mu^{2}A_{s}%
^{2})\alpha_{3}]Z$ and $\varphi_{s}=\varphi_{c}+\varphi=\omega_{s}T+[\left(
\mu^{2}A_{s}^{2}-\omega_{s}^{2}\right)  \alpha_{1}+\omega_{s}(3\mu^{2}%
A_{s}^{2}-\omega_{s}^{2})\alpha_{3}]Z$, which describes a bright soliton
solution with the maximal amplitude $A_{s}$. The other is that when the
soliton amplitude $A_{s}$ vanishes, it reduces to the CW light solution
$q_{c}(z,t)=Ae^{i\varphi_{c}}$. Therefore, in general, the exact solution
$q(z,t)$ in Eq. (\ref{S1}) describes a soliton solution embedded in a CW light
background with the group velocity $V_{sc}=(\nu_{1}M_{R}+\nu_{2}M_{I})/M_{I}$
\cite{lishuqing}. Specifically, when $\alpha_{3}=0$ and taking $\alpha_{1}%
=-1$, $\mu=1$, the solution (\ref{S1}) coincides with the result given in Ref.
\cite{Park}, which is firstly derived from NLS equation by N. Akhmediev and V.
I. Korneev \cite{Akhmediev8}.

\begin{figure}[ptb]
\centering\vspace{-0.0cm}
\includegraphics[width=8.2cm]{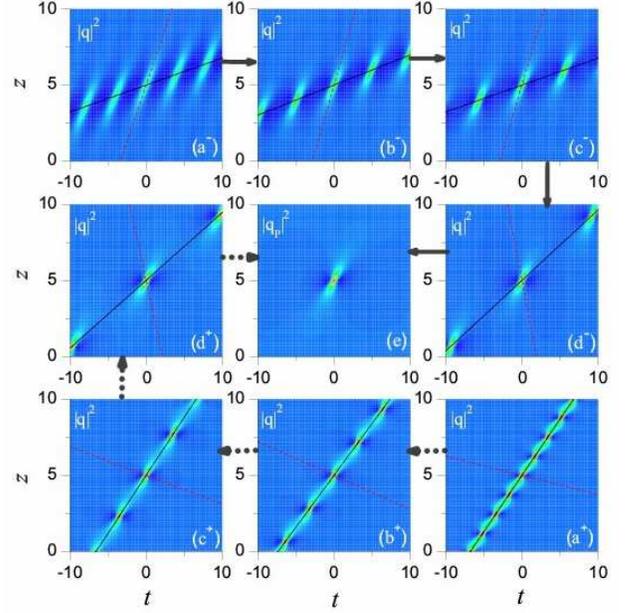}\caption{(Color online) The reduction
process from the solution (\ref{S1}) to the Peregrine solution
(\ref{Peregrine}) as $(A_{s},\omega_{s})\rightarrow(2A^{\mp},\omega)$.
(a$^{-}$) $A_{s}=A$, $\omega_{s}=0.6\omega$; (b$^{-}$) $A_{s}=1.45A$,
$\omega_{s}=0.7\omega$; (c$^{-}$) $A_{s}=1.6A$, $\omega_{s}=0.8\omega$;
(d$^{-}$) $A_{s}=1.99A$, $\omega_{s}=0.85\omega$; (a$^{+}$) $A_{s}=3A$,
$\omega_{s}=1.4\omega$; (b$^{+}$) $A_{s}=2.5A$, $\omega_{s}=1.3\omega$;
(c$^{+}$) $A_{s}=2.4A$, $\omega_{s}=1.2\omega$; (d$^{+}$) $A_{s}=2.01A$,
$\omega_{s}=1.15\omega$; and (e) the Peregrine solution given by
(\ref{Peregrine}) with $A_{s}=2A$ and $\omega_{s}=\omega$. The slope of red
dashed lines is $K_{\varphi}$, and black solid lines is $K_{\theta}$. Here the
parameters are $A=1$, $\omega=1$, $t_{0}=0$, $z_{0}=5$, $\mu=1$, $\alpha
_{1}=0.5$, and $\alpha_{3}=0.05$. }%
\end{figure}

In the limit $(A_{s},\omega_{s})\rightarrow(2A,\omega)$, Eq. (\ref{S1})
reduces to PS as follows%
\begin{equation}
q_{\text{p}}(z,t)=Ae^{i\varphi_{c}}\left[  \frac{4+i8CZ}{1+4C^{2}Z^{2}%
+4B^{2}\left(  T-DZ\right)  ^{2}}-1\right]  , \label{Peregrine}%
\end{equation}
where $B=\mu A$, $C=2B^{2}(\alpha_{1}+3\omega\alpha_{3})$ and $D=2\omega
\alpha_{1}+3(\omega^{2}-2B^{2})\alpha_{3}$, which is a rational fraction
solution, and is first derived from the Kuznetsov-Ma breather (KMB) by
Peregrine \cite{Peregrine,Kuznetsov,Ma} and so is called the Peregrine
solution or the Peregrine rogue wave (Note that it is regarded as the
Peregrine soliton in Refs. \cite{NatPhys,Hammani}.) Here $(t_{0},z_{0})$
presents the peak position of the PS. In order to understand the
characteristics of the solution (\ref{S1}), Fig. 1 presents the reduced
process from the solutions (\ref{S1}) to the PS (\ref{Peregrine}) as
$(A_{s},\omega_{s})\rightarrow(2A^{-},\omega)$ and $(A_{s},\omega
_{s})\rightarrow(2A^{+},\omega)$, respectively. From them one can see directly
that the solution (\ref{S1}) commonly exhibits breather characteristics, and
the separation between adjacent peaks gradually increases as $(A_{s}%
,\omega_{s})\rightarrow(2A,\omega)$, and eventually reduces into the PS. The
most impressive feature on the PS is localized in both time and space. The
peak position of the PS is fixed at spatiotemporal position $z=z_{0}$ and
$t=t_{0}$ during the reduction, and the peak power $|q_{\text{p}}(z_{0}%
,t_{0})|^{2}=(A+A_{s})^{2}\rightarrow9A^{2}$ as $A_{s}\rightarrow2A$, which
means that the PS with peak power $9A^{2}$ can be generated by choosing an
initial excitation properly.

It should be pointed out that in the reduction process mentioned above, the
physical mechanism for the formation of the PS has some differences. Figures
1(a$^{-}$)--1(e) demonstrate a localized process of CW background along the
slope direction $K_{\varphi}=M_{R}/(\nu_{2}M_{R}-\nu_{1}M_{I})$, while Figs.
1(a$^{+}$)--1(e) show a periodization process of bright soliton along the
slope direction $K_{\theta}=M_{I}/(\nu_{1}M_{R}+\nu_{2}M_{I})$. Furthermore,
Fig. 2 presents the contour plots of $K_{\theta}$ and $K_{\varphi}$ as a
function of $A_{s}$ and $\omega_{s}$ for given $A$ and $\omega$. Note that
$K_{\varphi}$ is infinite as $\nu_{2}M_{R}-\nu_{1}M_{I}=0$ except for the
point $(2A,\omega)$, which corresponds to the line $t=t_{0}$ and appears at
the sixth curve in Fig. 2(b). When $(A_{s},\omega_{s})\rightarrow(2A,\omega)$,
the limitations of $K_{\varphi}$ and $K_{\theta}$ do not exist, and at this
point the solution is localized along both slope directions $K_{\varphi}$ and
$K_{\theta}$, namely, the PS appears. Therefore, the PS should be a middle
state in the process of the localization of CW converting into the
periodization of the bright soliton, as shown in Ref. \cite{Akhmediev3}.

\begin{figure}[ptb]
\centering\vspace{-0.0cm}
\includegraphics[width=8.3cm]{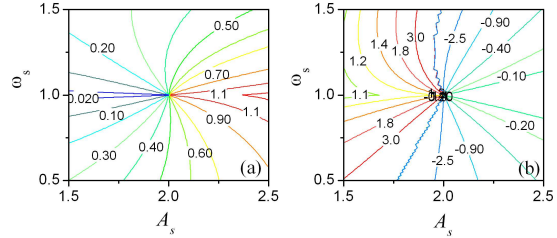}\caption{(Color online) The contour
plots of (a) $K_{\theta}$ and (b) $K_{\varphi}$ as a function of $A_{s}$ and
$\omega_{s}$, where the parameters are $A=1$, $\omega=1$, $\mu=1$, $\alpha
_{1}=0.5$, and $\alpha_{3}=0.05$, and the contour values (from left to right)
are $0.02,$ $0.1$, $0.2$, $0.3$, $0.4$, $0.5$, $0.6$, $0.7$, $0.9$, $1.1$,
$1.3$ in (a), and the contour values are $1.1$, $1.2$, $1.4$, $1.8$, $3$,
$-2.5$, $-0.9$, $-0.4$, $-0.2$, $-0.1$ in (b), respectively. }%
\end{figure}

In order to understand better the reduction processes of the PS, we consider a
special case for the solution (\ref{S1}) with $\omega_{s}=\omega$. In this
case, the solution (\ref{S1})\ has two different expressions. When $A_{s}%
^{2}<4A^{2}$, the solution (\ref{S1}) can be written as%
\begin{equation}
q(z,t)=\left(  \Omega_{R}\frac{\Omega_{R}\cosh\theta+iA_{s}\sinh\theta
}{2A\cosh\theta-A_{s}\cos\varphi}-A\right)  e^{i\varphi_{c}}, \label{S11}%
\end{equation}
where $\theta=\mu^{2}A_{s}\Omega_{R}(\alpha_{1}+3\omega\alpha_{3})Z$ and
$\varphi=\mu\Omega_{R}T-\mu\Omega_{R}[2\omega\alpha_{1}+(3\omega^{2}-2\mu
^{2}A^{2}-\mu^{2}A_{s}^{2})\alpha_{3}]Z$ with $\Omega_{R}=\sqrt{4A^{2}%
-A_{s}^{2}}$ being a modulation frequency. Here $K_{\theta}=0$ and
$K_{\varphi}=1/[2\omega\alpha_{1}+(3\omega^{2}-2\mu^{2}A^{2}-\mu^{2}A_{s}%
^{2})\alpha_{3}]$. Therefore the solution (\ref{S11}) is periodic with period
$t_{\text{per}}=2\pi/(\mu\Omega_{R})$ along the $t$ axis and localized along
the $z$ axis, and is usually called an Akhmediev breather (AB), which can be
considered as a modulation instability process \cite{Ablowitz}. In this case,
the solution (\ref{S11}) can reduce to the CW background $q_{c}(z,t)$ as
$A_{s}\rightarrow0$, and with the increasing of $A_{s}$ the CW background is
gradually localized due to the interaction with the soliton and forms a
periodic breather with period $t_{\text{per}}$, i. e, the AB [see Fig. 3(a)],
eventually when $A_{s}$ tends to $2A$, the solution (\ref{S11}) becomes the
PS. This process represents the reduction of CW$\rightarrow$AB$\rightarrow$PS.
Also, it can be described by the ratio between the period $t_{\text{per}}$ and
the temporal width $\delta t_{0}$ as follows:%
\begin{equation}
\frac{t_{\text{per}}}{\delta t_{0}}=\frac{2\pi}{\cos^{-1}\left(  \frac{A_{s}%
}{A}-2\frac{A}{A_{s}}\right)  }, \label{S13}%
\end{equation}
where the temporal width $\delta t_{0}$ is defined as the width from
zero-value of intensity to adjacent peak \cite{NatPhys}. Figure 3(b) presents
the dependence of the ratio $t_{\text{per}}/\delta t_{0}$ on $A_{s}/A$. From
it, one can see that when $A_{s}/A$ approaches $2$, the ratio $t_{\text{per}%
}/\delta t_{0}$ tends to infinity, which implies that the separation between
peaks is more and more large, eventually resulting in the localization along
the $t$ axis, and forms the PS. From the expression (\ref{S13}), one can see
that this process does not depend on the higher-order parameter $\alpha_{3}$,
which means that when the higher-order effects with the conditions $\alpha
_{2}=2\mu^{2}\alpha_{1}$, $\alpha_{4}=$ $6\mu^{2}\alpha_{3}$, and $\alpha
_{4}+\alpha_{5}=0$ simultaneously appear in the optical fiber, they do not
influence\ the characteristics of the PS. It should be emphasized that the
Peregrine solution generated by the process of the reduction of CW$\rightarrow
$AB$\rightarrow$PS has been already studied theoretically and experimentally
in the framework of the NLS equation
\cite{Osborne,Akhmediev3,Ma,Akhmediev8,NatPhys,Hammani}. Here we presented the
corresponding descriptions in the femtosecond regime.

\begin{figure}[ptb]
\centering\vspace{-0.0cm}
\includegraphics[width=8.3cm]{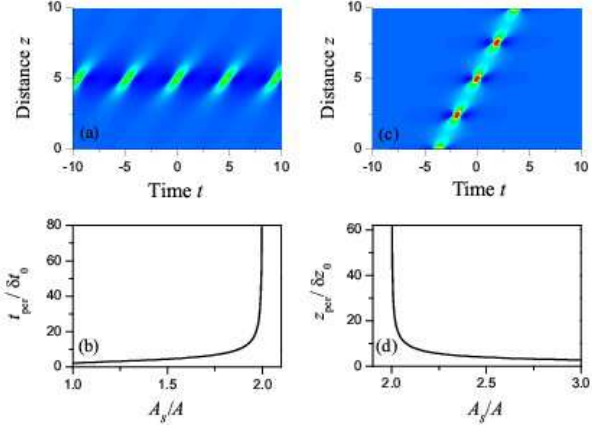}\caption{(Color online) (a) and (c) The
evolution plots of the solutions (\ref{S11}) and (\ref{S15}), respectively,
where the parameters are the same as in Fig. 1 except for $A_{s}=1.5A$ in (a)
and $A_{s}=2.5A$ in (c); (b) and (d) the dependences of the ratio
$t_{\text{per}}/\delta t_{0}$ and $z_{\text{per}}/\delta z_{0}$ on $A_{s}/A$,
respectively.}%
\end{figure}

When $A_{s}^{2}>4A^{2}$, the solution (\ref{S1}) can reduce to the following
form%
\begin{equation}
q(z,t)=\left(  \Omega_{I}\frac{\Omega_{I}\cos\varphi+iA_{s}\sin\varphi}%
{A_{s}\cosh\theta-2A\cos\varphi}-A\right)  e^{i\varphi_{c}}, \label{S15}%
\end{equation}
where $\theta=\mu\Omega_{I}T-\mu\Omega_{I}[2\omega\alpha_{1}+(3\omega^{2}%
-2\mu^{2}A^{2}-\mu^{2}A_{s}^{2})\alpha_{3}]Z$ and $\varphi=\mu^{2}A_{s}%
\Omega_{I}(\alpha_{1}+3\omega\alpha_{3})Z$ with $\Omega_{I}=\sqrt{A_{s}%
^{2}-4A^{2}}$. Here $K_{\theta}=1/[2\omega\alpha_{1}+(3\omega^{2}-2\mu
^{2}A^{2}-\mu^{2}A_{s}^{2})\alpha_{3}]$ and $K_{\varphi}=0$. Especially since
$A_{s}\gg2A$, namely $\Omega_{I}\sim A_{s}$, the solution (\ref{S15}) can be
approximated as $q(z,t)\approx q_{s}(z,t)-q_{c}(z,t)$, which is the
superposition of a CW solution and a bright soliton with the larger amplitude
$A_{s}$. From the solution (\ref{S15}), one can see that it is a periodic
function with period $z_{\text{per}}=2\pi/\mu^{2}A_{s}\Omega_{I}(\alpha
_{1}+3\omega\alpha_{3})$ along the $z$ axis and is localized along the $t$
axis, possessing the periodic peaking property\textit{ }of the field amplitude
like the Kuznetsov-Ma soliton (KMS) \cite{Kuznetsov,Ma,Xu}, and so is usually
called Kuznetsov-Ma soliton, as shown in Fig. 3(c). In this case, the solution
(\ref{S15}) can reduce to the bright soliton $q_{s}(z,t)$ as $A\rightarrow0$,
and with the increasing of $A$, the bright soliton is periodized due to the
interaction with the CW background and forms a KMS, and eventually reduces
into the PS. This process represents the reduction of bright
soliton$\rightarrow$KMS$\rightarrow$PS. Similarly, this process can be
described by the ratio between the period $z_{\text{per}}=2\pi/\mu^{2}%
A_{s}\Omega_{I}(\alpha_{1}+3\omega\alpha_{3})$ and $\delta z_{0}$ in the form%
\begin{equation}
\frac{z_{\text{per}}}{\delta z_{0}}=\frac{\pi}{\cos^{-1}\left[  \frac
{A_{s}(-7A^{2}-2AA_{s}+A_{s}^{2})}{2A(A^{2}-2AA_{s}-A_{s}^{2})}\right]  },
\label{S17}%
\end{equation}
where $\delta z_{0}$ is defined as the distance between two corresponding
locations of half of the peak intensity along the slope direction $K_{\theta}%
$. Figure 3(d) presents the dependence of the ratio $z_{\text{per}}/\delta
z_{0}$ on $A_{s}/A$. From it, one can see that when $A_{s}/A$ approaches $2$,
the ratio $z_{\text{per}}/\delta z_{0}$ tends to infinity, which means that
the distance between peaks is more and more large, eventually resulting in the
formation of the PS. Similarly, the expression (\ref{S17}) does not depend on
the parameter $\alpha_{3}$.

Another feature of PS can be described by the $z$-independent integral and the
energy exchange between the PS and the CW background. Indeed, by introducing
the light intensity against the CW background as follows,
\begin{align}
E_{c}(t,z)  &  =\left\vert q_{p}\left(  t,z\right)  \right\vert ^{2}%
-\left\vert q_{p}\left(  \pm\infty,z\right)  \right\vert ^{2}\nonumber\\
&  =A^{2}\frac{8+32C^{2}Z^{2}-32B^{2}\left(  T-DZ\right)  ^{2}}{[1+4C^{2}%
Z^{2}+4B^{2}\left(  T-DZ\right)  ^{2}]^{2}}, \label{density}%
\end{align}
it can be shown that it possesses the $z$-independent integral property, i. e.
$\int_{-\infty}^{+\infty}E_{c}(t,z)dt=0$. From the condition $E_{c}(t_{0}%
\pm1/2B,z_{0})=0$, one can define the width of PS as $1/B$, and have integrals
$\int_{-1/2B}^{+1/2B}E_{c}(t,z_{0})dt=4A^{2}/B$ and $\int_{-\infty}%
^{-1/2B}E_{c}(t,z_{0})dt+\int_{+1/2B}^{\infty}E_{c}(t,z_{0})dt=-4A^{2}/B$.
These results show that the energy of PS with stronger intensity mainly
concentrates in its central interval $(-1/2B,1/2B)$, but as a result of
$z$-independent integral properties, it loses the same energy in the
background so that the relation $S_{1}+S_{2}=S_{3}$ holds, as shown in Figs.
4(a)-4(c). It should be noted that because $B=\mu A$ is independent of the
higher-order terms in Eq. (\ref{HirotaEq}), the higher-order effects do not
influence this property on the PS.

The energy exchange between the PS and the CW background is of the form%
\begin{align}
E_{e}(z)  &  =\int_{-\infty}^{+\infty}\left\vert q_{p}\left(  t,z\right)
-q_{p}\left(  \pm\infty,z\right)  \right\vert ^{2}dt\nonumber\\
&  =\frac{4A^{2}\pi}{B\sqrt{1+4[2B^{2}(\alpha_{1}+3\omega\alpha_{3})]^{2}%
Z^{2}}}. \label{quantity}%
\end{align}
From the expression (\ref{quantity}), one can see that $E_{e}(z)$ is a
periodic function of $z$, which differs from the periodic exchange between the
bright soliton and the CW background in Eq. (\ref{S1}) \cite{lishuqing}. Also,
one finds that $E_{e}(z)$ is monotonously increasing as $z<z_{0}$ and is
monotonously decreasing as $z>z_{0}$; thus it takes a maximal value at
$z=z_{0}$, as shown in Fig. 4(d). This result shows that at $z=z_{0}$ the
energy exchange between the PS and the CW background reaches maximum. In order
to understand the influence of the higher-order effects on the PS, Fig. 4(d)
presents the evolution plots of $E_{e}(z)$ for different TOD parameter
$\alpha_{3}$. From it, one can see that the increase of the TOD parameters
$\alpha_{3}$ can enhance the rate of the energy exchange, as shown in Figs.
4(a) and 4(b).

\begin{figure}[ptb]
\centering\vspace{-0.0cm}\includegraphics[width=8.3cm]{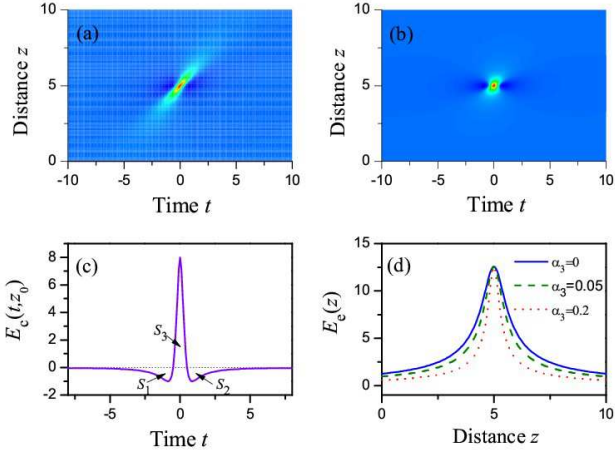}\vspace
{-0.0cm}\caption{(Color online) The evolution plots of the Peregrine solution
given by Eq. (\ref{Peregrine}) for $\alpha_{3}=0$ in (a) and $\alpha_{3}=0.2$
in (b), respectively. (c) The distribution of the light intensity against the
CW background given by Eq. (\ref{density}) at $z=z_{0}$. (d) The evolutions of
the energy exchange between the PS and the CW background for the different TOD
parameters. Here the parameters are $A=1$, $\omega=1$, $t_{0}=0$, $z_{0}=5$,
$\mu=1$, and $\alpha_{1}=0.5$.}%
\end{figure}

\section{Initial excitations of the Peregrine rogue waves}

In this section we will discuss the initial excitations of the Peregrine rogue
wave. We start with considering the excitation of the Peregrine rogue wave
based on the solutions (\ref{S11}) and (\ref{S15}). For solution (\ref{S11}),
by linearizing its initial expression, one finds that the initial expression
of the solution (\ref{S11}) can be approximated by
\begin{equation}
q_{\pm}(0,t)\approx\left(  \rho_{\pm}+\epsilon_{\pm}\cos\varphi\right)
e^{i\varphi_{c}(0,t)}, \label{s14}%
\end{equation}
where $\rho_{\pm}=(2A^{2}-A_{s}^{2}\mp iA_{s}\Omega_{R})/(2A)$ with
$\left\vert \rho_{\pm}\right\vert =A$, $\epsilon_{\pm}=A_{s}\Omega_{R}%
(\Omega_{R}\mp iA_{s})/(2A^{2})\exp[\mp\mu^{2}A_{s}(\alpha_{1}+3\omega
\alpha_{3})\Omega_{R}z_{0}]$, and $\varphi=\mu\Omega_{R}T+\mu\Omega
_{R}[2\omega\alpha_{1}+(3\omega^{2}-2\mu^{2}A^{2}-\mu^{2}A_{s}^{2})\alpha
_{3}]z_{0}$. Here the subscripts \textquotedblleft$+$\textquotedblright\ and
\textquotedblleft$-$\textquotedblright\ correspond to the case of $\alpha
_{1}+3\omega\alpha_{3}>0$ and $\alpha_{1}+3\omega\alpha_{3}<0$, respectively.
It can be shown that $\epsilon_{\pm}\rightarrow0$ as $A_{s}\rightarrow2A$.
Thus the expression (\ref{s14}) can be regarded as an initial condition with a
small periodic perturbation of background with the period $t_{\text{per}}$.
The numerical simulations show that the evolution of the exact solution
(\ref{S11}) can be well described by the initial approximation (\ref{s14})
(also see Ref. \cite{lishuqing}). Here one makes use of the initial
approximation (\ref{s14}) to excite the Peregrine rogue wave as $A_{s}$ closes
to $2A$.

\begin{figure}[ptb]
\centering\vspace{-0.0cm} \includegraphics[width=8.3cm]{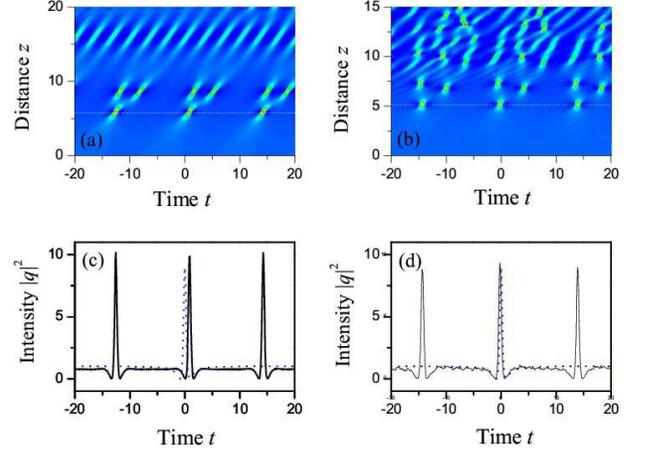}\vspace
{-0.0cm}\caption{(Color online) The evolution plots of the numerical solution
of Eq. (\ref{HirotaEq}) with the initial condition (\ref{s14}) and the
comparisons of the intensity profiles of numerical and exact results at peak
position, respectively: (a,c) $\alpha_{3}=0$, (b,d) $\alpha_{3}=0.2$. Note
that in (c) and (d), the blue dotted curves are the exact results given by Eq.
(\ref{Peregrine}) at $z=z_{0}$ and the black solid curves are the numerical
results at $z=5.9$ in (a) and $z=5.2$ in (b) [the white dotted lines in (a)
and (b)], respectively. Here $A_{s}=1.95$, and other parameters are the same
as in Fig. 4. }%
\end{figure}

Figure 5 presents the evolution plots of the numerical solution of Eq.
(\ref{HirotaEq}) with the initial condition (\ref{s14}) and the comparisons of
the intensity profile between numerical and exact Peregrine rogue waves at
peak position for $\alpha_{3}=0$ and $\alpha_{3}\neq0$, respectively. From it,
one can see that when $A_{s}$ approaches to $2A$, the initial condition
(\ref{s14})\ evolves into a string of near-ideal Peregrine rogue waves.
Although the theoretical results reveal that there is a Peregrine rogue wave
only in the limitation of $A_{s}\rightarrow2A$, in practice, this cannot be
implemented due to $\epsilon_{\pm}=0$ as $A_{s}=2A$. Furthermore, from Figs.
5(a) and 5(b), it can be seen that the evolution for long distances shows the
breakup of the Peregrine rogue wave, which implies that the Peregrine rogue
wave is unstable. Also, from Figs. 5(c) and 5(d), one can see that the
presence of the higher-order effects did not markedly influence the intensity
distribution of the Peregrine rogue wave at peak position, except for a
displacement of the peak position, but shortens the distance of energy
exchange, as shown in Figs. 5(a) and 5(b). This is in agreement with that
suggested in Fig. 4(d).

\begin{figure}[ptb]
\centering\vspace{-0.0cm}
\includegraphics[width=8.3cm]{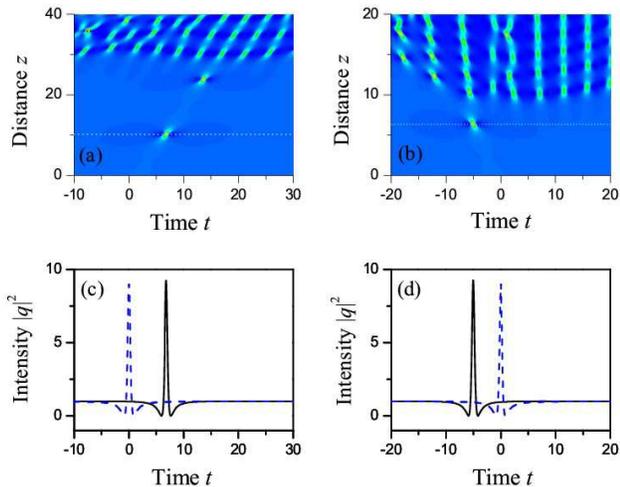}\caption{(Color online) The evolution
plots of the numerical solution of Eq. (\ref{HirotaEq}) with the initial
condition (\ref{S18}) and the comparisons of the corresponding intensity
profile at peak position, respectively: (a,c) $\alpha_{3}=0$, (b) and (d)
$\alpha_{3}=0.2$. Note that in (c,d), the blue dotted curves are the exact
results given by Eq. (\ref{Peregrine}) at $z=z_{0}$ and the black solid curves
are the numerical results at $z=10.2$ in (a) and $z=6.4$ in (b) [the white
dotted lines in (a) and (b)], respectively. Here $A_{s}=2.05$, $\omega=0.5$,
and other parameters are the same as in Fig. 4. }%
\end{figure}

In the following, we discuss the initial excitation induced by the solution
(\ref{S15}). In this case, one find that the solution (\ref{S15})\ can take
the following particular form%
\begin{align}
q(z_{1},t)  &  =\left(  -A+i\Omega_{I}\operatorname{sech}\theta\right)
e^{i\varphi_{c}(z_{1},t)}\nonumber\\
&  =Ae^{i\left[  \varphi_{c}(z_{1},t)+\pi\right]  }+\Omega_{I}%
\operatorname{sech}\theta e^{i\left[  \varphi_{c}(z_{1},t)+\frac{\pi}%
{2}\right]  } \label{S18}%
\end{align}
at the location $z_{1}=(2n\pi+\pi/2)/[\mu^{2}A_{s}\Omega_{I}(\alpha
_{1}+3\omega\alpha_{3})]+z_{0}$, $n=0,\pm1,\pm2,\cdots$, where $\theta
=\mu\Omega_{I}T-[2\omega\alpha_{1}+(3\omega^{2}-2\mu^{2}A^{2}-\mu^{2}A_{s}%
^{2})\alpha_{3}](2n\pi+\pi/2)/[\mu A_{s}(\alpha_{1}+3\omega\alpha_{3})]$.
Without loss of generality, here we take $n=0$. The expression (\ref{S18}) is
the superposition of a CW and an aperiodic hyperbolic secant function,
especially as $A_{s}\rightarrow2A$, $\Omega_{I}$ tends to zero. This means
that the expression (\ref{S18}) can be regarded as an initial condition with a
small aperiodic (simple peak) perturbation of background, which differs from
the superposition of a CW and a periodic perturbation in the expression
(\ref{s14}). Here we make use of the expression (\ref{S18}) as an initial
condition to investigate the excitation of the Peregrine rogue wave.
Similarly, the numerical simulations show that the evolution of the exact
solution (\ref{S15}) can be well described by the initial condition
(\ref{S18}) except for a translation.

Figure 6 shows the evolutions of the numerical solution of Eq. (\ref{HirotaEq}%
) with the initial condition (\ref{S18}) and the comparisons of the intensity
profile of numerical and exact Peregrine rogue waves at peak position for
$\alpha_{3}=0$ and $\alpha_{3}\neq0$, respectively. From it, one can see that
the initial condition (\ref{S18}) evolves into a near-ideal Peregrine rogue
wave, which differs from that shown in Figs. 5(a) and 5(b). Similarly, Figs.
6(a) and 6(b) show the breakup of the Peregrine rogue wave for long distance,
which implies that the Peregrine rogue wave is unstable.

\begin{figure}[ptb]
\centering\vspace{-0.0cm}
\includegraphics[width=8.3cm]{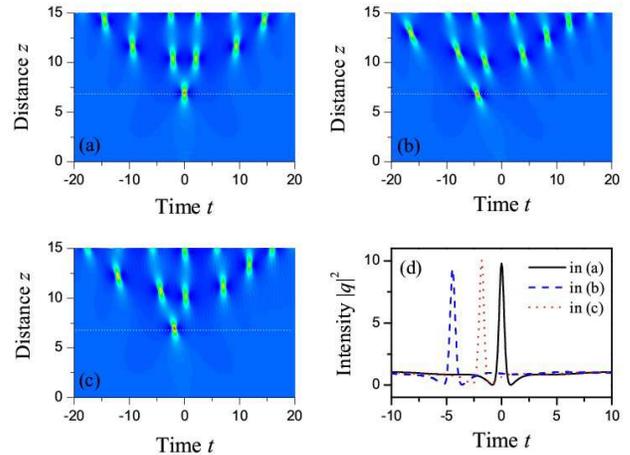}\caption{(Color online) The numerical
evolution plots of the initial condition (\ref{SS}): (a) for NLS equation with
$\alpha_{3}=0;$ (b) for Hirota equation (\ref{HirotaEq}) with $\alpha
_{3}=0.1;$ (c) for HNLS equation (\ref{He}) with $\alpha_{3}=0.05$,
$\alpha_{4}=0.285,$ and $\alpha_{5}=-0.315;$ and (d) the distributions of the
intensity profile at $z=6.975$ in (a), $z=6.825$ in (b), and $z=6.9$ in (c).
Here $\alpha_{1}=0.5$, $A=1$, $\epsilon=0.05,$ and $\sigma=0.05$. }%
\end{figure}

Comparing results in Figs. 5 and 6, we find that a small localized
(single-peak) perturbation pulse of CW background can excite a Peregrine rogue
wave. Indeed, every peak in the initial expression (\ref{s14}) can excite a
Peregrine rogue wave; thus a periodic perturbation of background resulted in
the generation of a string of the Peregrine rogue waves, as shown in Figs.
5(a) and 5(b). So we can suggest that the Peregrine rogue wave can be excited
by the interaction between the CW and a small localized (single-peak)
perturbation pulse. As an example, we consider a simple initial condition with
a Gaussian-type perturbation pulse as follows%
\begin{equation}
q(0,t)=A+\epsilon\exp(-\sigma t^{2}), \label{SS}%
\end{equation}
where $\epsilon$ is a modulation amplitude and a small quantity. The numerical
simulations show that when the width of the initial perturbation pulse is wide
enough, i. e, the parameter $\sigma$ is small enough, the Peregrine rogue wave
can be excited by the initial condition (\ref{SS}) even for a nonintegrable
HNLS equation (\ref{He}). Figure 7 shows the evolution plots of the numerical
solution for Eq. (\ref{He}) with the initial condition (\ref{SS}), which
includes a numerical evolution of the initial condition (\ref{SS}) under the
nonintegrable case, as shown in Fig. 7(c). From Fig. 7(d), one can see that
the main characteristics of the Peregrine rogue wave are maintained.
Certainly, the Peregrine rogue wave is unstable. This result can be used to
understand extreme localized events in ocean.

\section{The influence of the self-frequency shift to the dynamics of the
Peregrine rogue wave}

It should be noted that the equation (\ref{He}) does not include the
self-frequency shift effect arising from stimulated Raman scattering because
the parameter $\alpha_{5}$ is a real number. In this section we discuss the
influence of the self-frequency shift effect on the dynamics of the Peregrine
rogue wave. In this case, the model governing pulse propagation can be written
as \cite{Agrawal}
\begin{align}
\frac{\partial Q}{\partial\xi}  &  =-i\frac{\beta_{^{2}}}{2}\frac{\partial
^{2}Q}{\partial\tau^{2}}+i\gamma|Q|^{2}Q+\frac{\beta_{^{_{3}}}}{6}%
\frac{\partial^{3}Q}{\partial\tau^{3}}\nonumber\\
&  -\frac{\gamma}{\omega_{0}}\frac{\partial|Q|^{2}Q}{\partial\tau}-i\gamma
T_{R}Q\frac{\partial|Q|^{2}}{\partial\tau}, \label{model}%
\end{align}
where $\tau$ and $\xi$ represent the temporal coordinate and the propagation
distance, $\beta_{^{2}}$ is the group velocity dispersion, $\beta_{3}$ is the
third-order dispersion, $\gamma$ is the nonlinear coefficient of the fiber,
and $T_{R}$ is the Raman time constant. By introducing the dimensionless
transformations $Q(\xi,\tau)=\sqrt{P_{0}}q(z,t)$, $t=\tau/T_{0}$, and
$z=\xi/L_{D}$ with the dispersion length $L_{D}=T_{0}^{2}/|\beta_{^{2}}|$, Eq.
(\ref{model}) becomes the form of Eq. (\ref{He}) with the coefficients
$\alpha_{1}=-\beta_{^{2}}/(2\left\vert \beta_{^{2}}\right\vert )$, $\alpha
_{2}=\gamma L_{D}P_{0}$, $\alpha_{3}=\beta_{^{_{3}}}L_{D}/(6T_{0}^{3})$,
$\alpha_{4}=-\gamma L_{D}P_{0}/(\omega_{0}T_{0}),$ and $\alpha_{5}=-i\gamma
T_{R}L_{D}P_{0}/T_{0}$, respectively. It should be pointed out that here
$\alpha_{5}$ is a complex number which describes the self-frequency shift
effect arising from stimulated Raman scattering.

As an example, here we use realistic parameters for a highly nonlinear fiber
at $1550$nm with the group velocity dispersion $\beta_{^{2}}$ $=-8.85\times
10^{-1}$ ps$^{2}$/km, the third-order dispersion $\beta_{^{_{3}}}%
=1.331\times10^{-2}$ ps$^{3}$/km, and the nonlinear parameter $\gamma=10$
W$^{-1}\cdot$km$^{-1}$ \cite{NatPhys}. Thus, for a given initial power $P_{0}%
$, the parameters $\alpha_{1}$, $\alpha_{2}$, $\alpha_{3}$, $\alpha_{4}$ and
$\alpha_{5}$ can be determined. Note that in our simulations we take
$\alpha_{2}=1$ by choosing a temporal scale $T_{0}=[|\beta_{^{2}}|/(\gamma
P_{0})]^{1/2}$, and the Raman time constant $T_{R}=5$fs when the
self-frequency shift is considered. We still take Eq. (\ref{SS}) as the
initial condition, in which the realistic width of the initial perturbation
pulse is $T_{1}\equiv T_{0}/\sqrt{2\sigma}$.

Figure 8 presents the numerical evolution plots of the initial condition
(\ref{SS}) with the initial perturbation pulse width $T_{1}=2$ps for the
different initial power $P_{0}$. From Fig. 8 it can be seen that, in the
absence of the self-frequency shift effect ($T_{R}=0$), the Peregrine rogue
wave in turn splits into two subpulses, three subpulses, and so on, and can
split into more subpulses for higher initial power, as shown in Figs. 8(a) and
8(b). These results are similar to the pulse splitting induced by higher-order
modulation instability based on NLS equation in Ref. \cite{Erkintalo}, but
here the complex splitting dynamic evolutions are excited by a linear
superposition of the CW and a small localized (single peak) perturbation
pulse, and the higher-order effects, such as the third-order dispersion and
the self-steepening, are included. However, when the self-frequency shift
effect is considered, such splitting of the Peregrine rogue wave no longer
appears, as shown in Figs. 8(c) and 8(d). It is surprising that, in this case,
the dynamics of the Peregrine rogue wave mainly exhibit a peak changing
propagation characteristic\textit{ }of the field amplitude except for some of
the small radiations, and has an increase in peaking value, as shown in Figs.
8(e) and 8(f). This property can be used for generation of the higher peak
power pulse.

\begin{figure}[ptb]
\centering\vspace{0cm} \includegraphics[width=8.3cm]{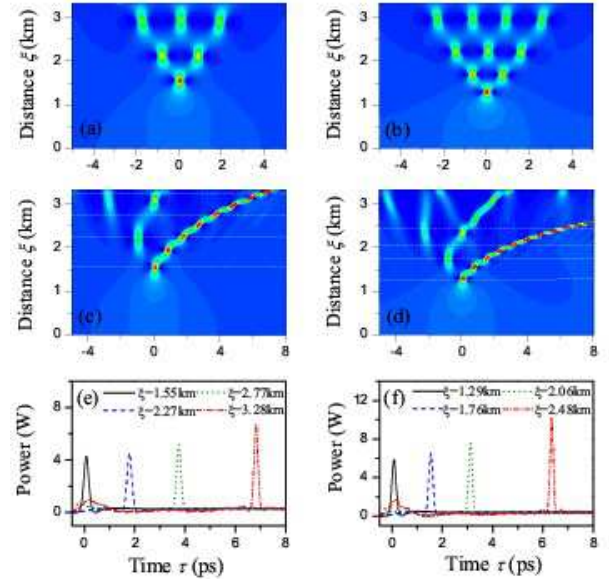}\caption{(Color
online) The numerical evolution plots of the initial condition (\ref{SS}) for
the different initial power: (a,b) the absence of self-frequency shift
($T_{R}=0$), (c,d) the presence of self-frequency shift ($T_{R}=5$ fs), and
(e,f) the corresponding power density distributions at the different distance
labeled by the white dotted lines in (c) and (d), respectively. In (a), (c),
and (e), the initial power $P_{0}=0.3$W, and in (b), (d), and (f), the initial
power $P_{0}=0.4$W. Here $\epsilon=0.2$ and $A=1$. }%
\end{figure}

Furthermore, the dependences of the peak position $\xi_{\text{peak}}$ of the
excited Peregrine rogue wave on the modulation amplitude $\epsilon$ are
considered, as shown in Fig. 9. From it one can see that for a given initial
perturbation pulse width $T_{1}$ or initial power $P_{0}$, the peak position
$\xi_{\text{peak}}$ of the excited Peregrine rogue wave is a decreasing
function of the modulation amplitude $\epsilon$. Thus one can control the
position of the excited Peregrine rogue wave by suitably choosing the initial
perturbation pulse width or the initial power. Also, we find that the
self-frequency shift effect does not influence the peak position
$\xi_{\text{peak}}$ of the excited Peregrine rogue wave.

\begin{figure}[ptb]
\centering\vspace{0cm} \includegraphics[width=8.3cm]{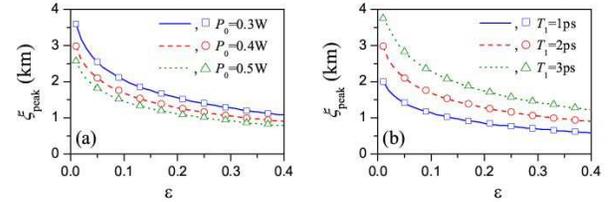}\caption{(Color
online) The dependences of the peak position $\xi_{\text{peak}}$ of the
excited Peregrine rogue wave on the modulation amplitude $\epsilon$ for (a)
the different initial power and a given initial perturbation pulse width
$T_{1}=2$ps, and (b) the different initial perturbation pulse width $T_{1}$
and a given initial power $P_{0}=0.4$W. Here the solid, dashed and dotted
curves correspond to the results of $T_{R}=0$, and the open diamonds, circles
and triangles correpond to the results of $T_{R}=5$fs, and the parameter
$A=1$. }%
\end{figure}

\section{Conclusions}

In summary, based on the soliton solution on a CW background for an integrable
Hirota equation, we have presented the reduction mechanism and the main
characteristics of the Peregrine rogue wave in the propagation of femtosecond
pulses of optical fiber. The results have shown that there exist two processes
of the formation of the Peregrine rogue wave: one is the reduction of
CW$\rightarrow$AB$\rightarrow$PS, which is the localized process of the CW
background, and the other is the reduction of bright soliton$\rightarrow
$KMS$\rightarrow$PS, which is the reduction process of the periodization of
the bright soliton. The characteristics of the Peregrine rogue wave have been
exhibited by strong temporal and spatial localization. Also, the initial
excitations of the Peregrine rogue wave have been discussed. The results have
shown that a Peregrine rogue wave can be excited by a small localized (single
peak) perturbation pulse of CW background, even for the nonintegrable HNLS
equation. This means that the Peregrine rogue wave is a result of interaction
between the continuous wave background and soliton. Furthermore, the numerical
simulations have shown that the Peregrine rogue wave is unstable. Finally,
through the study for a realistic highly nonlinear fiber, it has been found
that the self-frequency shift influences the dynamics of the Peregrine rogue
wave. The results show that in the absence of the self-frequency shift, the
Peregrine rogue wave can splits into several subpulses; however, when the
self-frequency shift is considered, the Peregrine rogue wave no longer splits
and exhibits mainly a peak changing and increasing propagation property.

This research was supported by the National Natural Science Foundation of
China under Grant No.61078079 and the Shanxi Scholarship Council of China
under Grant No.2011-010.

\end{document}